# A Brief Overview of Novel Approaches in Designing High Performance VCOs

Omid Reyhani-Galangashi
Department of Electrical, Biomedical and Mechatronics Engineering
Qazvin Branch, Islamic Azad University
Qazvin, Iran
reyhani@qiau.ac.ir

*Abstract*—Not only a voltage controlled oscillator (VCO) is one of the most significant component of every telecommunication system, but also it has been widely used in many other high-speed systems. In fact, a VCO has an important role in system operation, in other word, VCO is the heart of a system which gives existence to it. Nowadays, designing a high performance VCO for different applications is a challenging task for engineers. Moreover, up to now many designs and solutions have been proposed by the scientists to improve the performance of VCOs to be exploited in different cutting-edge applications. In this paper I will give a brief overview of the new techniques in designing high performance VCOs. Furthermore, the results of the proposed solutions have been compared with each other to see which method has advantages over the others.

## I. INTRODUCTION

In a high-speed system design, it is required that a phase-locked loop (PLL) works to add timing feature for clock control, data recovery, and synchronization [1]. In addition, a PLL in its simplest form is a feedback loop consisting of a phase detector and a voltage controlled oscillator (VCO); therefore, the VCO is one of the most important units of the PLL [2].

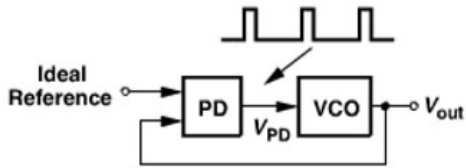

Fig. 1. A simple PLL block diagram. Source: Adapted from [2], copyright 2012, Prentice Hall.

As it is clear from VCO name, the output frequency of VCO is controlled by the variation of the input voltage level. Over and above that, in designing RF systems and especially in VCO, many parameters should be considered. The important parameters include, oscillation frequency, frequency tuning range, VCO gain which is measured in GHz/V or MHz/V, output power which is usually expressed in mW or dBm, phase noise (PN) in term of dBc/Hz and determined at 100KHz or 1MHz offset from the carrier frequency, and the temperature stability of output frequency which is measured in ppm/C [3].

This paper is organized in this way that in section II a short overview of the theoretical modeling of phase noise and tuning range is presented. In section III I will investigate the state-of-the-art designs which particularly aim at minimizing the VCO phase noise, maximizing the frequency tuning range as well as minimizing the power consumption. Finally, in Section IV I will conclude the paper.

## II. THEORITICAL MODELING

### A. Phase Noise

In the existing phase noise models, the power supply and the substrate noises have not usually been taken into account. However, in high-speed systems, power supply and substrate noises are higher than the device noise, and therefore for a precise design of a VCO, a detailed look into the phase noise analysis is mandatory [4]. According to Leeson's model, phase noise can be presented as follow [5]:

$$L(\Delta f) = 10 Log \left\{ \frac{2FkT}{P_s} \left[ 1 + \left(\frac{f_0}{2Q_L \Delta f}\right)^2 \right] \left[ 1 + \frac{f_{1/f^3}}{\Delta f} \right] \right\} \quad (1)$$

where F is noise factor, $k$ is the Boltzmann's constant, $T$ is the temperature, $P_s$ is the average consumed power in the tank circuit, $Q_L$ is the quality factor of the tank with all load elements included, $f_0$ is the oscillation frequency, $\Delta f$ is frequency offset and $f_1$ is the corner frequency.

In [6], a modification for Leeson's model is presented which shows that the Q factor has a direct impact on the phase noise of a VCO. The modified equation is shown in Eq. (2) which it can be seen that the phase noise is inversely proportional to the Q, thus increasing the Q factor will result in better phase performance.

$$L(\Delta \omega) = 10 Log \left[ \frac{2FkT}{P_s} \left(1 + \frac{\omega_0}{2Q \Delta \omega}\right)^2 \right] \quad (2)$$

Although Eq. (2) shows the relationship between Q and phase noise, a more detailed modification of Leeson's model is presented in [7], which is shown in Eq. (3).

$$L(f_m) = 10 Log \left\{ \left[ 1 + \left( \frac{f_0}{(2 f_m Q_L)\left(1 - \frac{Q_L}{Q_0}\right)} \right)^2 \right] \left(1 + \frac{f_c}{f_m}\right) \frac{FkT}{2P_{s,av}} + \frac{2kTRK_0^2}{f_m^2} \right\} \quad (3)$$

In Eq. (3), a new factor $f_c$, has been introduced which corresponds to the flicker frequency, $K_0$ is the modulation sensitivity expressed in (Hz/V) and $Q_0$ gives the unloaded quality factor of the tank circuit.

In [8] it is shown that the technology scaling also has impact on the phase noise and tuning range. Nowadays, in the manufacturing of VCO, a combination of both thin and thick gates transistor are used in a similar technology in order to simultaneously operate in both low and high voltage applications. By utilizing the thick gate transistors, it is possible to exploit higher supply voltages which lead to a lower thermal phase noise. On the other hand, improving the phase noise by using the mentioned method results in lower operating frequency, narrower tuning range, and higher power consumption.

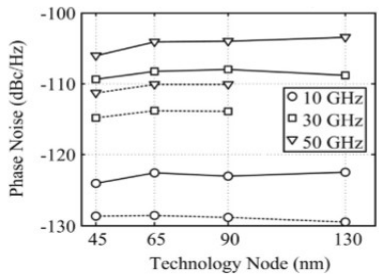

Fig.2. Technology trend effect on VCO phase noise at 1 MHz offset. (Solid lines: utilizing thin gate devices, dashed lines: utilizing thick gate devices). Source: Adapted from [8], copyright 2015, Elsevier.

*B. Tuning Range*

The majority of today's telecommunication applications require a tunable oscillator. In an ideal case, the output frequency of a VCO is a linear function of a control voltage or current inputs. However, having a good Frequency Tuning Range (FTR) is not achievable without degrading the phase noise or other parameters of the VCO.

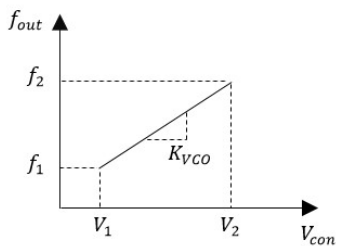

Fig.3. VCO gain or sensitivity diagram. Source: Adapted from [9], copyright 2016, IJSEAS.

As shown in [9], the simple definition for a VCO tuning range can be determined by Eq. (4). Additionally, the output frequency can be expressed by Eq. (5), in which $f_0$ is the center oscillation frequency, $K_{VCO}$ is the VCO gain and $V_{con}$ is the input voltage or control voltage.

$$FTR = f_2 - f_1 \quad (4)$$

$$f_{out} = f_0 + K_{VCO} \cdot V_{con} \quad (5)$$

Also it is worth to mention here that having a satisfying FTR in VCO is very important, due to covering the band width of the desired application or for compensating the misalignment of center frequency.

Usually tuning a VCO is implemented through exploiting the switched capacitor banks as shown in [10]. In [11] a novel method for tuning the frequency is proposed which employs the inductive tuning shown in Fig.4. which is an alternative method for conventional capacitive tuning. In this method, the FTR can be determined directly from Eq. (8) in which $k$ is coupling factor for loaded transformer.

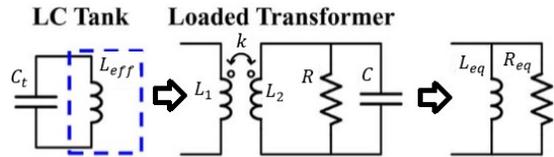

Fig.4. Inductive tuning of loaded transformer circuit model. Source: Adapted from [11], copyright 2014, IEEE.

$$\omega_1 = 1 \Big/ \sqrt{L_1 C_t} \quad (6)$$

$$\omega_2 = 1 \Big/ \sqrt{L_2 C} \quad (7)$$

$$FTR = \frac{\frac{\omega_1}{\sqrt{1-k^2}} - \omega_1}{\frac{\omega_1}{\sqrt{1-k^2}} + \omega_1} \times 2 = \frac{1-\sqrt{1-k^2}}{1+\sqrt{1-k^2}} \times 2 \quad (8)$$

It is clear from Eq. (8) that increasing coupling factor $k$ will result in FTR enhancement. Also, it should be considered that Eq. (9) is the optimum design condition for the proposed circuit presented in [11].

$$\omega_1 \ll \omega_2 \quad (9)$$

In [12], a varactor diode combined with multi-element array of resonant tunneling diode (RTD) VCO has been empirically proposed to increase the tuning range in Terahertz applications. In this method, the mechanism for frequency variation is based on the depletion layer of the varactor in which the maximum frequency is achieved when the varactor diode is entirely depleted and vice versa. Indeed, the output frequency is function of the bias voltage of varactor diode which can be controlled by changing the slot antenna width. In fact, slot antenna acts as a resonator and radiator as shown in Fig.5.

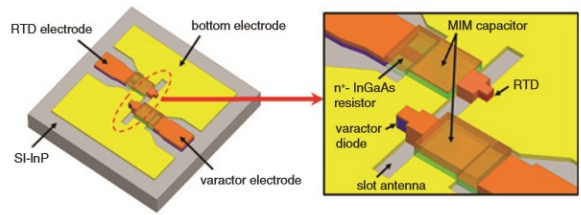

Fig.5. The RTD VCO combined with varactor diode for improving the tuning range in THz applications. Source: Adapted from [12], copyright 2016, IET.

### III. RECENT DESIGNS FOR HIGH PERFORMANCE VCOs

In this section an overview of the recent advances in the field of designing VCOs with low phase noise and enhanced FTR will be presented.

*A. A novel design using capacitive devision*

In [13], a VCO with boosted output swing and enhanced phase noise has been proposed using the Dynamic Threshold Metal Oxide Semiconductor (DTMOS) including capacitive division as shown in Fig.6.

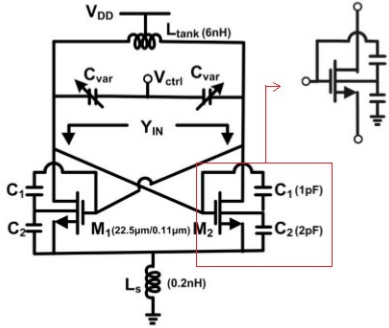

Fig.6. A novel VCO design using capacitive division technique. Source: Adapted from [13], copyright 2016, IEEE.

In existing DTMOS VCOs, the output swing should have a small value to keep the substrate-source P-N junction in the reverse biased mode. Conversely, the efficiency of the phase noise is proportional to the output swing, which means the larger the output swing, the better is the phase noise. Therefore, this capacitive division solution would drive the substrate which results in decreasing the threshold voltage in case of higher output swing [13].

$$V_{th} = V_{th0} + \gamma \left( \sqrt{\left|2\phi_F + \left(V_S - \frac{C_1}{C_1 + C_2} V_G\right)\right|} - \sqrt{|2\phi_F|} \right) \quad (10)$$

In Eq. (10) $V_{th0}$ is the initial threshold voltage, $\gamma$ is the body effect co-efficient, $\phi_F$ is related to substrate work function, $V_S$ is the source voltage and $V_G$ is the voltage of DTMOS gate.

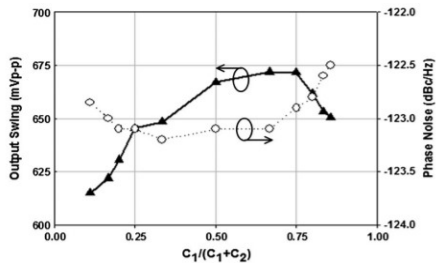

Fig.7. The optimized value for $C_1$ and $C_2$ ratio to have the optimum phase noise and output swing. Source: Adapted from [13], copyright 2016, IEEE.

In Fig.6. the tail inductor and the associated parasitic capacitance of the transistor's source, function as a noise filter which will omit the second harmonics and boost the swing amplitude [13], [14].

*B. A novel design combined with with class-B VCO*

In recent years many class-C VCOs have been proposed to have low power and improved phase noise VCO. The class-C VCO provides better phase noise performance than the other VCO structures. The mechanism for this structure has been investigated in [15]. The main reason for better noise performance of this class of VCO is due to the low voltage of the gate bias. As a result, a new class-C VCO has been proposed in [16], which has an auxiliary class-B pair that will help to boost the startup of the VCO. The schematic of the proposed design is shown in Fig.8.

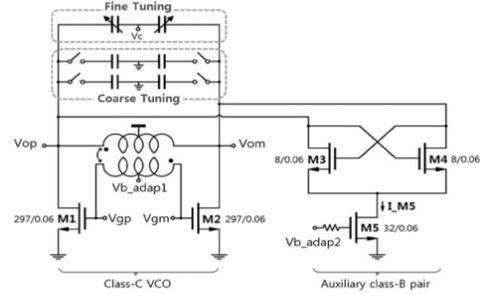

Fig.8. A class-C VCO paired with auxiliary class-B VCO for robust startup. Source: Adapted from [16], copyright 2016, IEEE.

The proposed circuit in [16] uses an adaptive circuit which causes the auxiliary part (Class-B) turns off automatically when the output signal goes to the steady state. This adaptive circuit is shown in Fig.9.

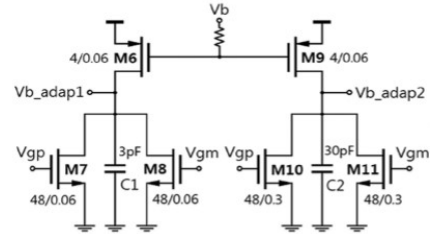

Fig.9. Adaptive biasing circuit. Source: Adapted from [16], copyright 2016, IEEE.

*C. A novel design using Self-Adjusted Active Resistor(SAAR)*

In [17], a high performance VCO is proposed. Therefore, as can be seen in Fig.10. a pair of PMOS transistors which are placed between LC tank and cross-coupled pair acts in the SAAR role which will get small resistance to itself when the cross-coupled transistors are in the saturation region. Smaller resistance values for SAAR will result in enabling sharp switching as well as abolishing the conversion of flicker noise to $1/f^3$ phase noise. Alternatively, SAAR will get larger resistance and prevent the small conducting resistance of the cross-coupled transistor degrade the quality factor of LC tank when cross-coupled transistor enters triode region [17]. The conducting resistance value of the $M_{p1}$ and $M_{p2}$ can be extracted out of Eq. (11).

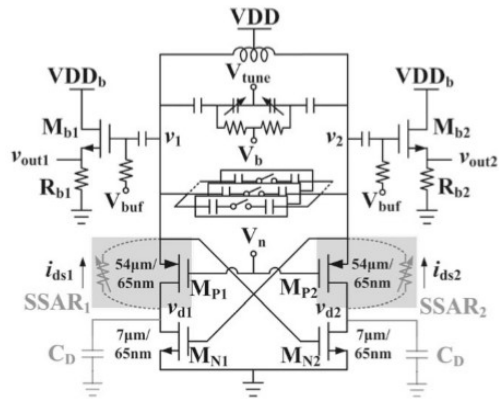

Fig.10. Schematic of a VCO using SAAR technique to achieve high performance in phase noise and power usage. Source: Adapted from [17], copyright 2016, IEEE.

$$r_{SAAR} = \frac{dv_{ds,p}}{di_{ds}} = \frac{L}{KW(|v_{gd,p}| - |V_{th,p}|)} \quad (11)$$

Now in Table I. a comparison between different parameters of the proposed VCOs can be seen.

Table I. Performance comparison of the proposed techniques.

| Ref. | Freq. (Ghz) | Technology | PN (dBc/Hz) @ 1 MHz | FTR |
|---|---|---|---|---|
| [17] | 5.71 | 65 nm | -113.7 | 22.4% |
| [16] | 2.46 | 65 nm | -132.41 | 10% |
| [13] | 2.31 | 0.11μm | -122 | 10.5% |

## IV. CONCLUSION

In this paper, I have introduced the VCO and its applications. Besides, I have presented the theoretical modeling for phase noise and FTR as well as the novel modifications for them. In addition, the impact of technology scaling on the phase noise of a VCO has been discussed briefly. Furthermore, as shown in section II of this paper, the RTD VCO combined with varactor diode which is proposed in [12] can enhance the frequency tuning range in THz applications. Also, in section III I have investigated three of the state-of-the-art VCO designs which were proposed recently. Eventually, as shown in Table I., the VCO design using the proposed method in [17], can bring higher FTR than the other methods at the expense of phase noise performance.